\documentclass[preprint]{aastex} 
\usepackage{epsfig}
\usepackage[caption=false]{subfig}
\usepackage{csquotes}
\usepackage{graphicx}
\usepackage{ulem,color}
\usepackage[dvipsnames]{xcolor/xcolor}
\usepackage{dcolumn}
\usepackage{bm}
\usepackage{natbib}
\usepackage[bookmarks=true]{hyperref}
\usepackage{xspace}
\usepackage{amsmath}
\usepackage{lipsum}
\usepackage{enumitem}

\setlist[itemize]{noitemsep}

\def\lesssim{\mathrel{\hbox{\rlap{\hbox{\lower4pt\hbox{$\sim$}}}\hbox{$<$}}}}
\def\gtrsim{\mathrel{\hbox{\rlap{\hbox{\lower4pt\hbox{$\sim$}}}\hbox{$>$}}}}

\slugcomment{Submitted to Ap.J.}

\begin{document}

\title{Quasi-Periodic Erupters: A Stellar Mass-Transfer Model for the Radiation}

\author{Julian H. Krolik$^{1}$ and Itai Linial$^{2}$}

\affil{$^1$William H. Miller Department of Physics and Astronomy, Johns Hopkins University, Baltimore MD 21218\\
  $^2$ Racah Institute of Physics, Hebrew University, Jerusalem, Israel}

\email{jhk@jhu.edu}

\begin{abstract}
Quasi-Periodic Erupters (QPEs) are a remarkable class of objects exhibiting very large amplitude quasi-periodic X-ray flares.  Although numerous dynamical models have been proposed to explain them, relatively little attention has been given to using the properties of their radiation to constrain their dynamics.  Here we show that the observed luminosity, spectrum, repetition period, duty cycle, and fluctuations in the latter two quantities point toward a model in which: a main sequence star on a moderately eccentric orbit around a supermassive black hole periodically transfers mass to the Roche lobe of the black hole; orbital dynamics lead to mildly-relativistic shocks near the black hole; and thermal X-rays at the observed temperature are emitted by the gas as it flows away from the shock.  Strong X-ray irradiation of the star by the flare itself augments the mass transfer, creates fluctuations in flare timing, and stirs turbulence in the stellar atmosphere that amplifies magnetic field to a level at which magnetic stresses can accelerate infall of the transferred mass toward the black hole.
\end{abstract}

\keywords{}

\section{Introduction}
\label{sec:introduction}

In galaxies hosting Quasi-Periodic Erupters (QPEs), bright X-ray flares close to the galaxies' centers repeat every few hours, raising the X-ray flux by as much as two orders of magnitude \citep{Miniutti+2019,Giustini+2020,Arcodia+2021,Arcodia+2022}.   At the peak of the flares, the luminosity reaches the range of X-ray luminosities seen in Seyfert galaxies, $\sim 10^{42} - 10^{43}$~erg~s$^{-1}$, but their spectra are much softer: roughly thermal with temperatures $\sim 100$~eV, rather than power-laws.  Although enough flares have been seen in a number of cases for the mean period to be fairly well-determined, in all cases the inter-flare intervals fluctuate by modest amounts, hence the adjective ``quasi-periodic".   Since the first example was published a few years ago, numerous mechanisms have been suggested to explain these enigmatic events: self-lensing of a pair of supermassive black holes \citep{Ingram+2021}, accretion disk instability \citep{Pan+2022}, interactions between stars and accretion disks \citep{Sukova+2021,Xian+2021}, and a number of variations on tidal stripping of nearby stars by supermassive black holes \citep{Metzger+2022,Wang+2022,Zhao+2022,King2022,LS2022}.

Despite the large amount of thought given to the dynamics that might create such events, there has been much less consideration of how to discriminate between proposed models using constraints based on the production of the X-rays.  The first goal of this paper is to separate viable from unviable scenarios by applying this method.  The results of this exercise are surprisingly specific, but they also uncover new questions: tidal stripping on its own cannot explain the duration of the flares, the size of the radiating area, or the fluctuations in inter-flare intervals.  The remainder of this paper seeks to answer these questions by making further inferences from the observed properties of these systems.  Most notably, a feedback loop can cause self-enhancement of the flare and introduce timing irregularities that may explain those seen in QPE lightcurves.

\section{Direct constraints}
\label{sec:constraints}

The phenomenology of QPEs places a number of very strong constraints on the underlying mechanism because we can measure many characteristic properties: a peak luminosity $L$; a color temperature $T$; an approximate period $P$; a flare duty cycle $D$, i.e., the ratio of flare duration to period; and the magnitude of variation in the time between two flares, i.e., the ``jitter".   No two events have exactly the same values of these quantities, but their range from one to the other is small enough that it makes sense to analyze a generic QPE in terms of a toy-model event possessing fiducial values for these parameters.   Here we choose $L_{\rm fid} = 5 \times 10^{42}$~erg~s$^{-1}$, $kT_{\rm fid} = 150$~eV (so that $T_{\rm fid} = 1.7 \times 10^6$~K), $P_{\rm fid} = 24000$~s, $D_{\rm fid}=0.2$, and typical jitter level $0.1P$.  In the following, when we derive values of other system properties, they can be scaled to the parameters characterizing any particular event in terms of ratios of these parameters to the fiducial values.  Such parameter ratios are denoted by script versions of the basic quantity: e.g., ${\cal L} \equiv L/L_{\rm fid}$.

\subsection{Size of the radiating region}

The X-ray spectra are generally well-fit by a thermal spectrum, at least in the sense that they decline as an exponential in energy in the observed band.  If their spectra are truly thermal, the luminosity and temperature taken together define a size scale for the radiating area.  If the shape of the radiation region is taken to be a two-sided thin circular disk, we find that this characteristic size is
\begin{equation}
R_{\rm th} = \left({L \over 2 \pi \sigma T^4}\right)^{1/2} = 3.9 \times 10^{10} {\cal L}^{1/2} {\cal T}^{-2}\hbox{~cm}.
\end{equation}
Thus, the characteristic scale of the radiating area is similar to the radius of a low-mass star.  Posed in terms of relativistic objects, it is roughly the gravitational radius $r_g \equiv GM/c^2$ for a black hole of mass $2.7 \times 10^5 M_\odot$.  Thus, if the event is associated with a black hole of mass $ \gtrsim 10^5 M_\odot$, its emitting region cannot resemble an accretion disk---the radiating area is too small.

It is important to recognize, however, that there may be other relevant lengthscales associated with the radiating region.  For example, although the condition of thermal radiation constrains the {\it surface area} of the region, it does not constrain its {\it thickness}.   If the region is less symmetric than a filled circular disk, $R_{\rm th}$ represents the geometric mean of the two orthogonal lengths characterizing the surface area, but one could be larger and the other smaller.  For example, one might suppose that the radiating material is in an accretion disk around a supermassive black hole, but occupying only a narrow annulus.  This might be possible in principle, but it would be challenging to create a dynamical picture yielding such a situation.   Lensing alters geometry, but it creates flares by magnifying the image, effectively {\it increasing} the radiating area.

There is also another caveat in applying this constraint.  It is possible the exponential decline of the spectrum is a signature of temperature, but for a non-Planckian spectrum, e.g., that of optically thin bremsstrahlung.  In this case, the luminosity and temperature combine to yield not the radiating surface area, but the emission measure \cite{RL}:
\begin{equation}
EM = 1.7 \times 10^{66} {\cal L} {\cal T}^{-1/2}\hbox{~cm$^{-3}$}.
\end{equation} 
However, for the region to be optically thin demands that its effective optical depth $(\tau_{\rm ff} \tau_T)^{1/2} < 1$, where $\tau_{\rm ff}$ is the free-free absorption optical depth and $\tau_T$ is the optical depth to Thomson scattering.  Its geometric thickness must then be greater than
\begin{equation}
h_{\rm brems} \approx 4 \times 10^{11} (\varepsilon/3)^{-6/5} {\cal L}^{3/5} {\cal T}^{-17/10}\hbox{~cm},
\end{equation}
where $\varepsilon \equiv h\nu/kT$.  Taken in ratio to $R_{\rm th}$, we find
\begin{equation}
h_{\rm brems}/R_{\rm th} \simeq 10 {\cal L}^{1/10} {\cal T}^{3/10}.
\end{equation}
In other words, to reproduce the luminosity and temperature of QPEs with optically thin bremsstrahlung, the thickness of the region must be at least an order of magnitude greater than the characteristic scale of its radiating area.  Although possible, such an oddly shaped region immediately suggests that optically thin bremsstrahlung is a somewhat less plausible model for the origin of the ``thermal" X-ray radiation of QPEs.  At the very least, it is even less compatible with a conventional accretion disk.

\subsection{QPE period as an orbit}

The intervals between flares in QPEs are sufficiently regular that there must be some sort of underlying stable clock.  In astrophysical systems, stable clocks are most often found in the form of either the rotation of some coherent body or a gravitational orbit.  Here we will assume the latter.

From this point of view, the characteristic orbital period $P_{\rm fid} = 24000$~s has a special significance: it is only a few times larger than the characteristic dynamical time of a star.   If a star is in an orbit with such a period, it immediately follows that the tidal gravity due to its partner in that orbit is a significant influence on the star.

The period also determines a semimajor axis for the orbit in terms of the total system mass \citep{Zhao+2022}:
\begin{equation}
a = 1.2 \times 10^{13} {\cal P}^{2/3} {\cal M}_{\rm BH}^{1/3}\hbox{~cm} = 80 {\cal P}^{2/3} {\cal M}_{\rm BH}^{-2/3} r_g.
\end{equation}
Here we have selected a fiducial partner mass $M_{\rm BH,fid} = 10^6 M_\odot$.  Thus, for any supermassive black hole partner, the orbital separation is close to the region where relativistic dynamics become important.  Relativistic dynamics mean that the energy per unit mass potentially available for radiation is large; it also means that relativistic orbital effects such as apsidal precession could be significant.

Because tidal gravity must be important, the most natural way for the orbit to function as the system's underlying clock is for the orbit to be eccentric, and the star loses mass near pericenter as a result of the stronger tidal gravity there.  To be more quantitative, the distance from the center of the star to the L1 point is
\begin{eqnarray}
R_{\rm L1} &\approx& 8.3 \times 10^{10} f(\phi,e) {\cal P}^{2/3} (10^6 M_*/M_{\rm BH})^{1/3} \hbox{~cm} \nonumber \\
&\approx& 1.2 R_*  \, f(\phi,e) {\cal P}^{2/3} {\cal M}_*^{-0.55} {\cal M}_{\rm BH}^{-1/3},
\end{eqnarray}
where $f(\phi,e)a$ is the distance from the star to the black hole as a function of azimuthal angle when the orbit has eccentricity $e$.  For the stellar mass $M_*$, we choose a fiducial value of $M_\odot$, and we adopt a main-sequence mass-radius relation in which $R_* \propto M_*^{0.88}$ \citep{Ryu2020b}.  Here we have also assumed that the star rotates synchronously with the orbit at the azimuthal angle of interest.  Although this is almost impossible to achieve because the angular velocity of an eccentric orbit continually changes around the orbit, it is a reasonable approximation because the value of $R_{L1}$ changes by only a fraction $\sim O(0.1)$ for order-unity changes in rotation rate \citep{Dai+Blandford}.  This estimate vindicates our assertion about the importance of tidal gravity: the distance from the stellar photosphere to the L1 point is only $\sim O(0.1) R_*$ at all points around the orbit, and (momentarily disregarding the star's response) that distance may shrink to zero--or even less--near the pericenter even for $e \sim O(0.1)$.  Thus, eccentric orbital periods of this scale around a supermassive black hole are almost guaranteed to lead to periodically-modulated tidal mass-loss.

In addition, because the ratio $q \equiv M_*/M_{\rm BH} \ll 1$, the difference in potential between the L1 and L2 points is quite small: $\sim q (GM_{\rm BH}/(fa)$ \citep{MD2000}.  Thus, in this situation mass-loss through L2 occurs at almost the same time as mass-loss through L1, although in general at a slightly lower rate \citep{LS2017}.

We can therefore expect mass-loss from the star for some period of time near pericenter passage, with essentially all of the mass lost through the L1 point captured by the black hole, as well as possibly some portion of the mass lost through the L2 point.  That the duty cycle $D \sim 0.2$ limits the range of eccentricities.  If the eccentricity is too small, $D$ would be too large.  If the eccentricity is too large, either $D$ would be very small, or so much of the star would overhang its Roche lobe when the star is near pericenter that a large part of its mass might be lost in a single orbit.  When classical Roche lobe analysis applies (circular orbit, corotating star), $D \propto [(1-e^3)/2e]^{1/2} (1 - R_{\rm L1}/R_*)^{1/2}$.  However, this scaling is valid only for $(1 - R_{\rm L1}/R_*) \ll e$ and becomes questionable for eccentric orbits because the star's atmosphere may not be able to respond quickly enough to the changing tidal gravity for this expression to apply.  In addition, as we shall see later in Sec.~\ref{sec:irradiation}, in these circumstances there may be strong transitory perturbations to the star's atmosphere due to effects other than tidal gravity.
Consequently, on qualitative grounds, we expect the combination of requiring the Roche lobe size to be at least $\sim 10\%$ smaller at pericenter than at a separation equal to the semimajor axis and the system to have a duty cycle $\sim 0.1$ limits the range of eccentricity to $0.1 \lesssim e \lesssim 0.5$.

\subsection{Flare energy and duration}

The total radiated energy per flare provides another constraint.  If the energy is derived from matter falling to a distance $R_{\rm diss}$ from the black hole, the mass whose free-fall energy must be dissipated in order to create a single flare is
\begin{equation}
\Delta M \sim 1 \times 10^{-7} {\cal L} {\cal D} {\cal P} (R_{\rm diss}/10 r_g) M_\odot.
\end{equation}
Note that this translates to a mass-loss rate $\sim 1 \times 10^{-4} {\cal L} {\cal D} (R_{\rm diss}/10r_g) M_\odot$~yr$^{-1}$.

All this energy must be radiated during the flare; if it were radiated over a longer period, the system would remain luminous during the inter-flare time.  In other words, the cooling time of the source region must be shorter than the flare duration. If the radiation region is roughly spherical, the length scale of the radiating area is also the region's thickness.  Assuming that all the mass delivered in a single episode is held in the radiating region, the condition for the cooling time to be shorter than the flare duration translates to
\begin{equation}
\frac{\kappa \Delta M}{2\pi R_{\rm th} c} < D P
\end{equation}
for opacity $\kappa$.  Using our estimate for $\Delta M$, this constraint becomes
\begin{equation}
{\cal L}^{1/2} {\cal T}^2 (R_{\rm diss}/R_{\rm th}) (\kappa/\kappa_T) \lesssim 0.3.
\end{equation}
Because the gas is certainly thoroughly ionized, $\kappa$ must be at least as great as the Thomson opacity $\kappa_T$.  However, we already know that $R_{\rm th} \simeq 0.3 {\cal L}^{1/2} {\cal T}^{-2} {\cal M}_{\rm BH}^{-1} r_g$, while $R_{\rm diss}$ must be $> r_g$, so that $R_{\rm diss}/R_{\rm th} \gtrsim 3 {\cal L}^{-1/2}{\cal T}^2 {\cal M}_{\rm BH}$.  It is therefore very difficult to satisfy this cooling time constraint unless ${\cal M}_{\rm BH} \lesssim 0.1$.
Because the host galaxies of known QPE events are not dwarfs, it is unlikely that their nuclear black hole is small enough.  It therefore follows that either the mass whose energy is used does not stay in the radiating region for the entire flare and/or the shape of the radiation region is far from spherical.

In addition, once the flare has ended, unless the radiating matter changes state sufficiently to produce unobservable photons (e.g. EUV), it cannot emit any substantial amount of additional energy---the inter-flare period is much fainter than the flare.  This statement is tantamount to saying that the matter cannot remain outside the black hole and suffer further dissipation for a time longer than the flare duration.  Because the free-fall time from $R_{\rm diss}$, $\sim 150 (R_{\rm diss}/10 r_g)^{3/2} {\cal M}_{\rm BH}$~s, is shorter than the fiducial flare duration $\sim 5000$~s, such a stipulation is plausible.

Lastly, the variation in flare duration and inter-flare interval raises another point: something about the structure of the star's outer layers must change from orbit to orbit.  If the star's structure were either stationary or changing on the timescale at which the star's mass changes, orbit-to-orbit variation in either of these timescales could not be as large as the observed $\sim 10\%$ level.

\subsection{Questions left open after application of direct constraints}

Although observed properties provide significant constraints on the system generating these events, they also raise several significant problems for models in which a star's orbit regulates the inter-flare intervals.  The scale of the radiating area is very small, too small to be understood in terms of accretion disk mechanics around supermassive black holes.  The flare duration is so short that a quasi-spherical source region containing all of a flare's mass would not cool quickly enough.
Lastly, the ``jitter" in the inter-flare intervals and their durations demands sizable fluctuations in the outer layers of the star on the timescale of the orbital period or shorter.

There is also another problem that has so far been left implicit: how the matter taken from the star falls to a depth $\sim R_{\rm diss}$.  So long as $q \ll 1$, the center-of-mass of the system lies very close to the black hole.  Consequently, the transferred matter's initial specific angular momentum with respect to the black hole is very close to the star's specific orbital angular momentum.   For our fiducial estimates, this angular momentum is, in units of $r_g c$,
\begin{equation}
J = 9 (1-e^2)^{1/2} {\cal P}^{1/3}{\cal M}_{\rm BH}^{-1/3}.
\end{equation}

This value is a few times larger than the specific angular momentum of an ISCO orbit around a non-spinning black hole, $\approx 3.5$, or the minimum angular momentum for a highly eccentric orbit that does not result in immediate capture, $4$ (the exact number for a non-spinning black hole, the approximate number for a spinning black hole after averaging over orbital orientation: \cite{Kesden2012}).   Consequently, if this gas retains its specific orbital angular momentum, and its radial speed (relative to the black hole) is small compared to its azimuthal speed, its pericenter cannot be very far inside the pericenter of the star's orbit.  Small radial speed is to be expected because the sound speed of gas in a stellar atmosphere is limiated by the escape velocity of the star, and this is smaller than the orbital speed by a factor $\sim (M_*/M_{\rm BH})^{1/3}$ when the distance from the black hole is similar to the tidal radius. Transfer of additional matter at the next pericenter passage would lead to shocks, but relatively weak ones.  Thus, the question of how the angular momentum of this gas can be reduced quickly enough for its pericenter to be reduced by at least a factor of order unity is central to the plausibility of this sort of model.

\section{Trajectory of mass transfer}

\subsection{Angular momentum transport by magnetic stresses}

In view of its importance to angular momentum transport in more conventional accretion flows (Balbus \& Hawley 1998), it is worth considering what contribution embedded magnetic fields might make to the solution of this problem.   We have already remarked that fluctuations in flare timing imply non-stationarity in the star's outer layers.  In the following section we suggest a mechanism to create these fluctuations.  For now, we will simply assume they are present.
A potential consequence of strong, irregular pressure fluctuations is turbulence.  A large body of work on turbulent dynamos \citep{Batchelor1950,Kulsrud1992,Scheko+2004,Seta2021,Bott+2021} has shown that large-amplitude (transonic) turbulence can amplify the magnetic field enough to make the Alfven speed $v_A \sim c_s$, for $c_s$ the local sound speed (in conditions of high opacity like a star, the sound speed reflects the combination of gas and radiation pressure).   As the gas being transferred accelerates away from the star (pulled by the black hole's tidal gravity and pushed by its internal pressure gradient), one might expect $v_A$ to increase above $c_s$ as the gas density drops while the component of the magnetic field parallel to the flow is unchanged.

\begin{figure}\label{fig:schematic}
\hskip 4cm \includegraphics[width=0.5\textwidth]{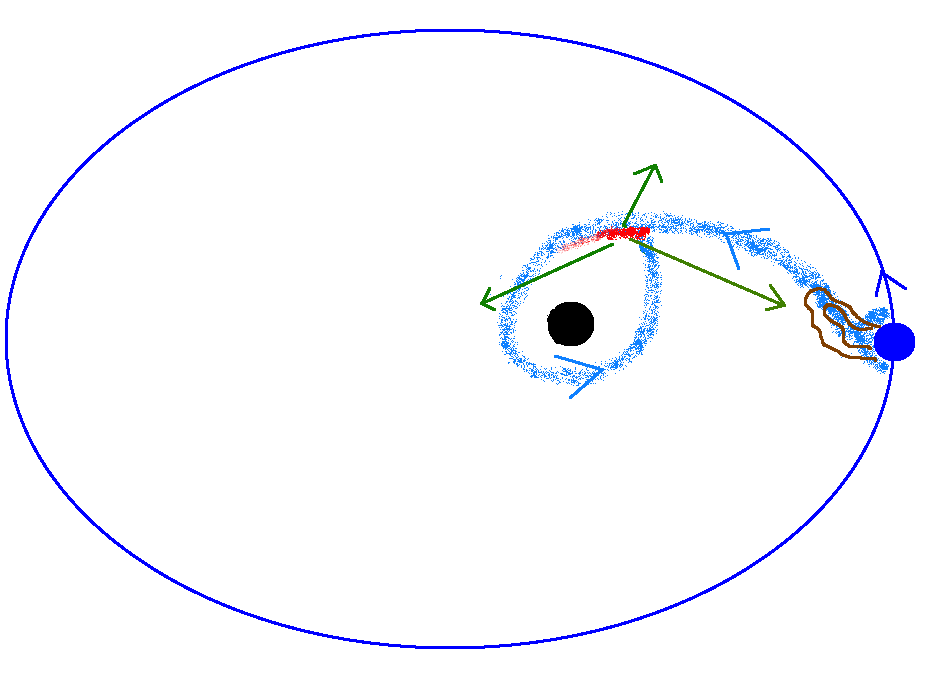}
\caption{Schematic diagram of mass-transfer.   The star (blue disk) follows an eccentric orbit around the black hole (black disk).   During the flare, its atmosphere expands outward on the side facing the black hole (dense blue shading; see Sec.~\ref{sec:irradiation}).   Gas from the atmosphere (lighter blue shading) streams through the L1 point, threaded by magnetic field (brown curves).   After losing angular momentum, the gas follows a more eccentric orbit that takes it close enough to the black hole to suffer sizable apsidal precession.  It shocks (dense red shading) against later-arriving gas and emits X-rays (green arrows).   Deflected post-shock gas cools rapidly  (fainter red shading).}
\end{figure}

Once the gas begins to orbit around the black hole, significant magnetic stresses should arise as orbital shear stretches the radial component of the field into the azimuthal direction (see Fig.~\ref{fig:schematic}).  If the gas travels within an annulus of width $\sim \chi R_*$ ($\chi \ll 1$ for gas expelled from a star through an L1 point), the timescale for the magnetic stress to remove a significant fraction of the gas's angular momentum is
\begin{equation}
t_{\rm torque} \sim \rho r v_\phi \left[\frac{\partial}{\partial r} \left( \frac{r B_r B_\phi}{4\pi}\right)\right]^{-1} \sim {\chi v_{\phi} R_* \over v_A^2}
\end{equation}
for gas density $\rho$ and speed perpendicular to the black hole direction $v_\phi$.
Phrased another way, the requirement for $t_{\rm torque}$ to be less than the flare duration is
\begin{equation}\label{eq:torquecondition}
 (R_*/r_g) (\chi v_\phi/v_A)^2  \lesssim 100 {\cal D} {\cal P}^{2/3} {\cal M}_{\rm BH}^{-2/3}.
\end{equation}
In other words, if $v_A \gtrsim 0.1 \chi v_\phi$, the magnetic field is strong enough to reduce the transferred matter's angular momentum in the time allowed.

It is important to note that a corollary of this mechanism for angular momentum transport is that whatever angular momentum is removed from the gas must immediately be given to something else.  Additional angular momentum given to the star causes the stellar orbit to expand.  The competition between orbital expansion and stellar expansion, whether provoked by mass-loss or external heating (which can be important to QPEs: see Sec.~\ref{sec:irradiation}) determines whether mass-loss grows or decays.  On the other hand, angular momentum given to gas that ultimately escapes the system has no effect on the star's orbit.

Rapid removal of angular momentum by magnetic stresses (or some other means) would also help settle several of the other open questions.   For example, when gas loses much of its angular momentum but little of its energy, the relation between pericenter and angular momentum (in Schwarzschild spacetime; dependence on spin parameter is fairly weak) becomes
\begin{equation}
r_p/r_g = (j^2/4)\left(1 - 16/j^2\right)^{1/2},
\end{equation}
where $j$ is the specific angular momentum in units of $r_g c$.  The apsidal precession per pericenter passage is then
\begin{equation}
\Delta\phi_p = \frac{40/j^2}{1 + \left(1 - 16/j^2\right)^{1/2}} \hbox{~rad}.
\end{equation}
If the gas's angular momentum were to quickly drop by as little as $\simeq 30\%$ (from $j \simeq 9$ to $j \simeq 6$), the apsidal precession per pericenter passage would be $\simeq 0.6$~rad.
Apsidal precession by angles of a radian or so creates strong shocks, as gas that has just passed through pericenter encounters infalling gas.  These shocks have large enough stream-stream velocity differences to dissipate amounts of energy not much smaller than the local gravitational potential (see Fig.~\ref{fig:schematic}).

Such shocks have a further benefit in this context: matter deflected inward with still lower angular momentum ($j \leq 4$) plunges directly into the black hole, shutting off any further radiation.

\subsection{The thermal and radiative implications of inter-stream shocks}

These remarks can be made more specific by examining the physical character of the shocks.  We begin by checking that the post-shock temperature of the gas is the one observed, i.e., $\sim 10^6$~K.  Because the upstream gas is cold, its pressure is likely gas-dominated, giving it an adiabatic index of 5/3.  On the other hand, the gas is so much hotter after passing through the shock that in LTE its pressure is overwhelmingly due to radiation; it therefore has an adiabatic index of 4/3.  In this sense, the shock is ``radiation mediated" \citep{LevinsonNakar2020}, but unlike many such shocks, the directions of both the shock propagation and the post-shock flow are in general not toward the photosphere. The high Mach number limit is an extremely good approximation here, so the immediate post-shock temperature is given by
\begin{equation}
aT^4 = \frac{3\gamma_2}{\gamma_1 (\gamma_2 +1)} \rho_1 v_s^2 = \frac{36}{35} \rho_1 v_s^2,
\end{equation}
where the subscripts 1 and 2 refer to pre- and post-shock, respectively, and $v_s$ is the shock speed.  

The density of the stream prior to the shock may be estimated from the flow rate required to generate the luminosity:
\begin{eqnarray}\label{eq:geomdens}
\rho_1 &=& \frac{L}{\pi R_*^2 c^3} \left(R_{\rm diss}/r_g\right)^{3/2} \nonumber \\
       &=& 4 \times 10^{-10} {\cal L} (R_*/R_\odot)^{-2} \left(R_{\rm diss}/10r_g\right)^{3/2} \hbox{~gm~cm$^{-3}.$}
\end{eqnarray}
Here we have estimated the cross section of the stream to be the size of the star; because this may be an overestimate of the cross section, the density could be higher. If the shock speed is roughly the virial speed, the post-shock temperature is
\begin{equation}
T \simeq 1.5 \times 10^6 \, \psi^{1/2} {\cal L}^{1/4} (R_*/R_\odot)^{-1/2} \left(R_{\rm diss}/10r_g\right)^{1/8} \hbox{~K}.
\end{equation}
The factor $\psi$ is the ratio of the stream width in units of $R_*$ to the shock speed in units of the virial speed; because both terms in the ratio are likely to be somewhat less than unity, this ratio is likely to be somewhere near unity, while the post-shock temperature depends only on its square root.
Thus the dynamics of this model---i.e., one in which matter taken from a star falls close to a black hole and suffers a shock with speed comparable to its pericentric orbital speed---automatically reproduce very closely the observed temperature.  In addition, the temperature's weak dependence on parameters is also consistent with the small spread of temperatures seen in these events.

\begin{figure}\label{fig:hotspot}
\hskip 4cm \includegraphics[width=0.5\textwidth]{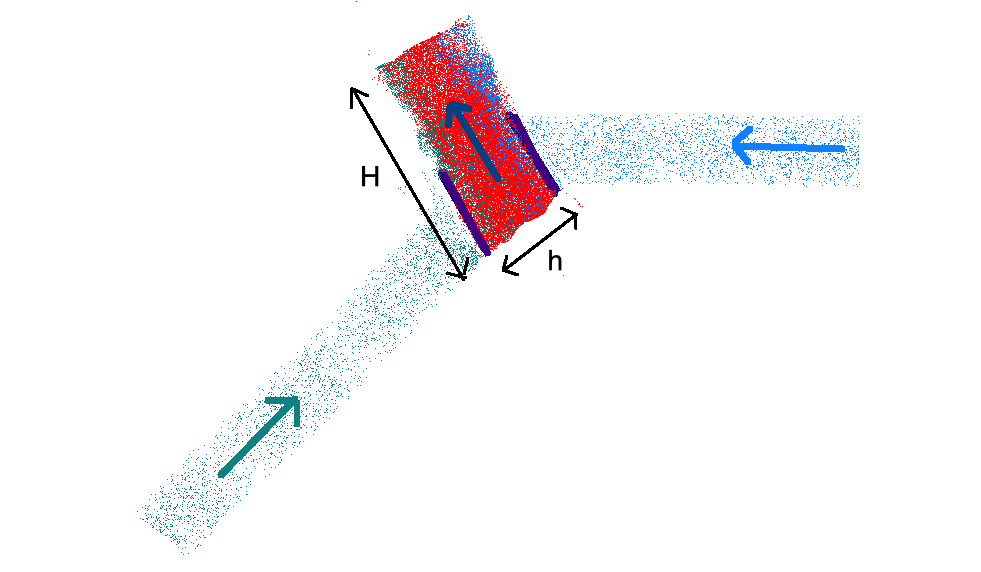}
\caption{Zoomed-in schematic diagram of the shocks and radiation region, showing only gas participating in the shocks.  Newly-arriving matter (blue shading and arrow) encounters matter whose orbit has been apsidally-rotated (green shading and arrow).  A pair of oblique shocks form (purple lines).  Shock-heated gas (red shading) follows a deflected trajectory while radiating and cooling.  Its width $h$ and cooling length $H \equiv \xi v_s t_{\rm cool}$ are shown with double-headed arrows.}
\end{figure}

The shocks also naturally create a source region of the sort suggested by our earlier estimate of the cooling time: one in which the gas passes through the source region, cooling fast enough that only a fraction of the transferred mass resides in the region at any given time, and the source region is far from spherical (see Fig.~\ref{fig:hotspot}).  In principle, the characteristic length scales of the post-shock flow in the two directions perpendicular to its velocity could be different from one another, particularly if the shock is oblique.  For the present purposes, we will ignore that possibility in the interest of simplicity.  The radiating area is then
\begin{eqnarray}
A_{\rm rad} &\sim& \xi v_s t_{\rm cool} h = 7 \phi v_s \kappa \rho_1 h^3/c \nonumber \\
            &\simeq& 1 \times 10^{23} \phi (\kappa/\kappa_T){\cal L} (h/R_*)^3 (R_*/R_\odot) \left(R_{\rm diss}/10r_g\right) \hbox{~cm$^2$}.
\end{eqnarray}
where $h$ is the single transverse length scale, and the post-shock fluid speed, including the component tangent to the shock front, is $\xi v_s$.  The factor 7 represents the high Mach-number shock compression factor for $\gamma = 4/3$.
The fiducial estimate for $A_{\rm rad}$ is $\sim 10\times$ the inferred radiating area, but a factor of 2 focusing of the streams, so that $h/R_* \sim 1/2$, would bring $A_{\rm rad}$ into agreement with the value inferred from observations.

As promised, the cooling time of the post-shock material also satisfies the constraint of being shorter than the flare duration:
\begin{equation}
t_{\rm cool} = 7\kappa \rho_1 h^2/c \simeq 200 (\kappa/\kappa_T) {\cal L} (h/R_*)^2 \left(R_{\rm diss}/10r_g\right)^{3/2} \hbox{~s}.
\end{equation}
If, as appears to be required in order to match the radiating area, $h/R_* \sim 1/2$, this time would become even shorter.

The sound-speed of the post-shock gas is
\begin{equation}
c_s = \left[\frac{(4/9) aT^4}{7\rho_1}\right]^{1/2} \simeq 2.5 \times 10^9 (R_{\rm diss}/10r_g)^{-1/2}\hbox{~cm/s}.
\end{equation}
Although likely blocked in some directions, in other directions the heated post-shock gas could possibly expand freely, growing at roughly $c_s$.  If so, in the time before it cools, its extent in that direction would grow to
\begin{equation}
h^\prime \sim 5 \times 10^{11} (\kappa/\kappa_T) {\cal L} (h/R_*)^2 (R_{\rm diss}/10r_g)^{1/2}\hbox{~cm}.
\end{equation}
Adiabatic cooling decelerates this expansion once $h^\prime \gtrsim h$, but only weakly because $c_s \propto \rho^{1/6}$.  If initially $h/R_* \sim 1/2$, the width of the radiating region would grow by a factor of a few over a cooling time, so that it is never larger (in width or length) than a few $r_g$ (see Fig.~\ref{fig:schematic}).
Even if it does grow, the radiating region should remain well-thermalized because the Planck mean opacity in these conditions is $\gtrsim 0.1 \kappa_T$ \citep{Hirose+2014}, and the column density is
\begin{eqnarray}
\Sigma &\sim& 7 \rho_1 h^\prime \nonumber \\
&\sim& 1 \times 10^3  {\cal L}^2 (\kappa/\kappa_T) (h/R_*)^2 (R_*/R_\odot)^{-2} \left(R_{\rm diss}/10r_g\right)^2 \hbox{~gm~cm$^{-2}$}.
\end{eqnarray}
Moreover, because the temperature of thermal emission for a given luminosity is so insensitive to radiating area, any change in spectral shape due to cooling, whether due to adiabatic expansion or radiation, might not be easily detectable.

When the matter whose orbit has suffered apsidal rotation shocks against newly-arriving matter, the sense of the deflection is to transfer angular momentum from the apsidally-rotated matter to the incoming matter.  As a result, the pericenter of the apsidally-rotated matter is reduced, possibly enough for it to be directly captured upon its next return to pericenter.   Such direct capture has the virtue of removing matter rapidly, thereby ensuring it cannot continue to radiate after the current episode of mass-transfer has ceased.

\section{Flare heating}
\label{sec:irradiation}

None of the preceding arguments accounts for either the turbulence invoked to explain the presence of strong magnetic field in the gas peeled away from the star or the fluctuations in inter-flare intervals and individual flare durations and luminosities.   Both strong turbulence in the star's outer layers and fluctuations in mass-transfer can be explained by a hitherto unrecognized effect that is inescapable in these circumstances: the intense irradiation of the side of the star facing the black hole by the X-rays emitted during the flare.  At pericenter, the ratio of the flare flux to the star's intrinsic flux is
$$F_X/F_* = 5 \times 10^4 (1-e)^{-2} {\cal L} {\cal P}^{-4/3} {\cal M}_{\rm BH}^{-2/3} (T_*/T_\odot)^{-4}.$$
If the star rotates synchronously with the orbit, this energy is always delivered to the same hemisphere of the star; if it rotates at some other rate, a given angular location on the star is periodically heated with a frequency that is the beat frequency between the stellar rotation rate and the orbital frequency.

In either case, such extremely strong heating must dramatically alter the structure of the star's atmosphere.  The energy fluence striking the star per orbit is enough to unbind a surface mass density
\begin{equation}
\Delta \Sigma_{\rm unbound} \sim 7 \times 10^3 (1-e)^{-2} {\cal L}{\cal D}{\cal P}^{-1/3} {\cal M}_{\rm BH}^{-2/3} {\cal M}_*^{-0.12}\hbox{~gm~cm$^{-2}$}
\end{equation}
from the exposed side of the star.  This is an upper bound on the amount of mass that could be liberated by heating because it assumes the energy is spread uniformly over a certain amount of mass and there are no radiative losses.   Integrating over the exposed stellar surface area suggests a maximum total unbound mass
\begin{equation}
\Delta M_{\rm unbound} \sim 5 \times 10^{-8} (1-e)^{-2} {\cal L}{\cal D}{\cal P}^{-1/3} {\cal M}_{\rm BH}^{-2/3} {\cal M}_*^{1.64} M_\odot.
\end{equation}
In other words, X-ray heating could account for an amount of mass removed from the star not much less than the amount required to explain the X-ray energy radiated.  It could therefore  significantly amplify the mass taken from the star and delivered to the black hole's Roche lobe, increasing it well above what might be expected from gravitational dynamics alone.

Because the heating agent is photons with energy $\sim 100$~eV, the temperature of the gas is unlikely to rise past $\sim 10^6$~K.  This, alone, is sufficient to drive substantial expansion (e.g., it is the temperature of the Solar corona), but the total pressure of the gas is rather greater than what would be expected on the basis of an ideal gas equation because the thermal energy density in this intensely-heated gas is radiation dominated.  For a gas density comparable to that of the Solar photosphere ($\approx 2 \times 10^{-7}$~gm~cm$^{-3}$), the LTE ratio of radiation pressure to gas pressure is $p_r/p_g = 450 {\cal T}^3$.  

Radiative losses are, in fact, likely to be at a noticeable level, but may not be strong enough to reduce $\Delta M_{\rm unbound}$ by a large factor.  If, after heating, the outer layer of the star has a scale height comparable to the stellar radius, its cooling time is comparable to the flare duration:
\begin{equation}
t_{\rm cool,atm} \sim 7 \times 10^3 (1-e)^{-2} {\cal L}{\cal D}{\cal P}^{-1/3} {\cal M}_{\rm BH}^{-2/3} {\cal M}_*^{0.76} \hbox{s}.
\end{equation}
Here, as would be appropriate to strongly heated gas, we assume the opacity is Thomson.

The fact that the cooling time when the gas has expanded to the stellar radial scale is comparable to the flare duration also implies that roughly this much mass will, in fact, be heated.  At a gas temperature $\sim 10^4$~K, collisional ionization equilibrium leads to an opacity for 200~eV photons $\sim 5 \times 10^{-21}$~cm$^2$~H$^{-1}$ (Krolik 1999).  Consequently, the impinging X-rays would initially be absorbed in a layer with surface density only $\sim 2 \times 10^{-4}$~gm~cm$^{-2}$, a very small fraction of $\Sigma_{\rm unbound}$.  However, having absorbed all the X-ray energy, this thin layer would rise dramatically in temperature, rapidly reducing its soft X-ray opacity.  As successively deeper layers are heated and their opacity is reduced to the Thomson level, the X-rays can penetrate deeper and deeper.  Precisely because the diffusion time through a layer with surface density $\sim \Delta \Sigma_{\rm unbound}$, Thomson opacity, and a thickness $\sim R_*$ is roughly the flare duration, the ionization wave would just barely reach the depth required by the end of the flare.

One of many possible dynamical effects due to such a dramatic break in the spherical symmetry of the star's pressure contours would be to create very strong fluid motions {\it around} the star as well as outward.  If these can persist for a few times the flare duration, the disorder they create could explain the unevenness in inter-flare intervals.  Thus, the X-ray heating of the star could inject the necessary fluctuations in flare timing that orbital mechanics alone would not produce.

These disordered motions would also powerfully stir turbulence in the star's outer layers.  If the heating is strong enough to unbind the gas, the speed of the large-scale eddies is comparable to the escape speed.  If, as is commonly the case in turbulent dynamos \citep{Batchelor1950,Kulsrud1992,Scheko+2004,Seta2021,Bott+2021}, the magnetic field energy density is amplified to rough equipartition with the turbulent kinetic energy density, the Alfven speed would then be comparable to the star's escape speed.  On the main sequence, this is almost always $\simeq 0.002c$, or $\sim 2 \times 10^{-2} v_{\rm orb}$ when $a \sim 100 r_g$.   The condition for rapid angular momentum transport posed by Equation~\ref{eq:torquecondition} demands an Alfven speed $\approx 5\chi$ larger.  If the density of the gas drops by a factor of several as it leaves the star's extended atmosphere and begins to orbit through the black hole's Roche lobe, and if $\chi$ (the ratio of stream-width to stellar radius) is also a factor of a few less than unity, this criterion could be satisfied.

\section{Summary}

Thus, the measured attributes of QPEs lead to a picture that reproduces {\it all} of their observed properties: the peak luminosity, the spectral color temperature, the mean interval between flares, the duty cycle of the flares, and even the departures from periodicity in the flares.   All can be understood in terms of a star that has arrived at a rather close orbit around a supermassive black hole ($a \sim 100 r_g$) with modest eccentricity; such a condition is not unexpected because a plausible dynamical pathway to such a state has been identified by \cite{LS2022}.  Moreover, the back-reaction of the luminous X-ray flares upon the star whose mass fuels the flares plays an important role.  It may create the turbulence in the star's atmosphere that accounts for the irregularities of the flares; in so doing, it can also increase the total amount of mass transferred in each pericenter passage.  In addition, by driving a turbulent dynamo, the strong X-ray heating creates the magnetic field under whose influence the mass falls quickly toward the black hole.

\section*{Acknowledgments}
We would like to thank Selma de Mink, Scott Noble, and Alberto Sesana for a conversation that stimulated this work.  Tsvi Piran and Re'em Sari helpfully commented on this manuscript at several stages during its preparation.  IL acknowledges support from the Adams Fellowship.  JHK  thanks the Kavli Institute for Theoretical Physics for providing hospitality and a venue for the initiating conversation; through KITP, this research was supported in part by the National Science Foundation under NSF grant PHY-1748958.  JHK was also partially supported by NSF grant AST-2009260.

\hfill \break
\bibliography{qpe.bib}

\end{document}